\documentclass[aps,showpacs,prx,amssymb, amsmath, twocolumn]{revtex4-2}
\usepackage{graphicx}
\usepackage{amssymb}
\usepackage{amsmath}
\usepackage{bm}
\usepackage{xcolor} 
\usepackage{array}
\usepackage{multirow}
\usepackage{tabularx}
\usepackage{booktabs}
\usepackage{textcomp}
\usepackage{mathtools}
\usepackage{bbold}
\usepackage{physics}
\usepackage{hyperref}

\begin{document}
\title{Programming higher-order interactions of Rydberg atoms} 
\author{Andrew Byun, Seokho Jeong, and Jaewook Ahn} \email{jwahn@kaist.ac.kr}
\address{Department of Physics, KAIST, Daejeon 34141, Korea}

\begin{abstract}
Higher-order interactions in spin-based Hamiltonians are crucial in addressing numerous fundamentally significant physical problems. In this work, Rydberg-atom graph gadgets are introduced to effectively program $K$-th order interactions within a Rydberg atom system. This approach facilitates the determination of the ground states of an Ising-type Hamiltonian, encoded to solve higher-order unconstrained optimization problems. A favorable scaling behavior, $O(N^K)$, is expected in terms of  the number of atoms required for $N$-vertex hypergraph optimization problems.
\end{abstract} 
\maketitle
\section{Introduction} 
Higher-order interactions in spin-based Hamiltonians play an important role in many fundamental physics problems. Exotic quantum phenomena, such as the Efimov trimer~\cite{Efimov1970, Naidon2017_Efimov}, fractional quantum Hall states~\cite{Cooper2004_FQH}, and topologically ordered states~\cite{Levin2005_stringnet, Levin2005_stringnet2} are attributed to non-binary spin interactions. For example, adding a three-body interaction term to the Hubbard model can lead to exotic phases with unique filling factors~\cite{Buchler2007_3polar, Schmidt2008_3B2Dlattice, BCS2009_3B1Dlattice, Bonnes2009_3B1Dlattice}. Moreover, higher-order interactions are also essential for molecular interactions involving more than four-body interactions due to their electronic structures~\cite{Seeley2012_H2} and for various high-energy physics models~\cite{Hauke2013_Schwinger, Pedersen2021_LGT, Farrell2023_QCD}.  In quantum information science, creating multi-qubit entangled states often requires many-body interactions beyond simple two-body interactions~\cite{Rossi2013_hyper, Liu2022_hyper}. Currently, topological phases are gaining attention for their potential in quantum error correction~\cite{Kitaev2003,Paetznick2013}, necessitating the incorporation of higher-order interactions in artificial quantum matter as well as in quantum information and computation~\cite{Bluvstein2022_Toric, Google2023, Bluvstein2024_logical, Self2024, Iqbal2024}.

In the Rydberg atom system~\cite{Browaeys2020}, two-body correlations naturally arise from the Rydberg blockade effect~\cite{Jaksch2000_blockade, Lukin2001_blockade, UrbanNatPhys2009_blockade, GaetanNatPhys2009_blockade}. However, implementing controllable many-body correlations presents experimental challenges, requiring complex energy level structures and precise electromagnetic field drivings~\cite{Glaetzle2017_LHZ, Gambetta2020_3Ryd, Pohl2009_AB, Faoro2015_Forster, Ryabtsev2018_Forster, Gurian2012_Forster, NEMJ2022_fractal}. Previous research has developed non-local two-body interactions using Rydberg quantum wires~\cite{Qiu2020, Kim2022_wire, Byun2022PRXQ_PlatonicSolid}, which incorporate an additional chain of atoms. By integrating Rydberg quantum wires with a three-dimensional (3D) configuration of atoms~\cite{Lee2016_3Drearrange, Barredo2018_3D, Kim2020_3DRyd, Song2021_Cayleytree}, all-to-all interactions between arbitrary pairs of atoms have been achieved~\cite{Qiu2020, Kim2022_wire, Kim2020_3DRyd}. Building upon the concept of the Rydberg quantum wire, which corresponds to a linear qubit graph, we aim to demonstrate the implementation of higher-order interactions. We introduce the design of new atomic qubit graphs representing the {\textit {hyperedge}} of a {\textit{hypergraph}}, thereby facilitating the representation of $K$-body correlations. Just as a system incorporating Rydberg quantum wires forms a Rydberg atom graph, representing both unweighted~\cite{Kim2022_wire, Byun2022PRXQ_PlatonicSolid} and weighted graph structures~\cite{Byun2023}, a system incorporating Rydberg hyperedges could represent a hypergraph.

The generation of hypergraphs are a scientific and technological challenge. Even for classical Ising models, conventional computational methods often prove inefficient~\cite{Liu2017_4B}. When higher-order interactions are involved, optimizing hypergraphs—which represent hyperedges corresponding to these interactions in the spin system—requires solving numerous NP (nondeterministic polynomial)-hard problems, such as the Max-$K$ SAT (satisfiability) problem for $K \ge 3$. Consequently, this method of generating higher-order interactions within the Rydberg atom graph could provide a viable solution for addressing both classical and quantum problems.

\begin{figure*}[t!]
\includegraphics[width=2.0\columnwidth]{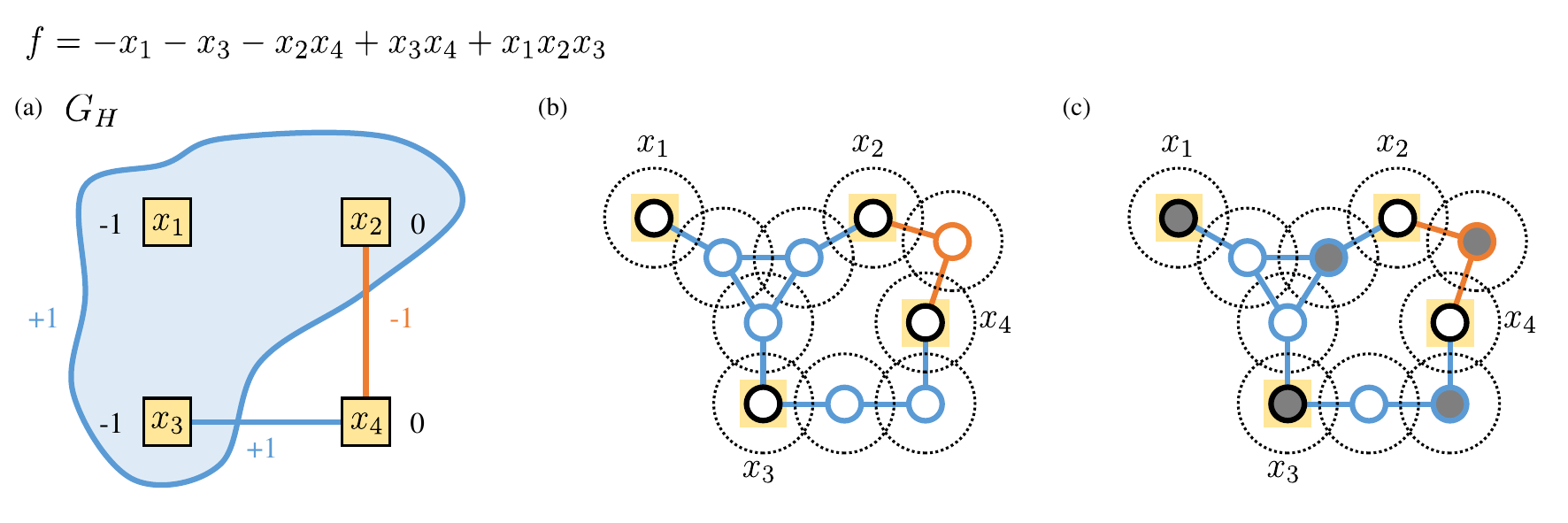}
\caption{(a) The hypergraph $G_H(V,E)$ representing the cost function $f=-x_1-x_3-x_2x_4+x_3x_4+x_1x_2x_3$. (b) The Rydberg atom graph constructed for the HUBO problem to optimize $f$. (c) The many-body ground state of the Rydberg atom system resulting in the HUBO solution, $(x_1,x_2,x_3,x_4)=(1,0,1,0)$.}
\label{Fig1}
\end{figure*}

As a conceptual overview of this paper, Fig.~\ref{Fig1} illustrates an example of the Rydberg-atom hypergraph representation of higher-order unconstrained binary optimization (HUBO). In the given example, the hypergraph $G_H(V, E)$ consists of four vertices, $V = \{x_1, x_2, x_3, x_4\}$, and three edges, $E = \{(x_2, x_4), (x_3, x_4); (x_1, x_2, x_3)\}$, where $(x_1, x_2, x_3)$ is an order-three ($K=3$) hyperedge, as shown in Fig.~\ref{Fig1}(a). This hypergraph $G_H$ represents the hypergraph optimization problem with the cost function $f = -x_1-x_3-x_2x_4 + x_3x_4 + x_1x_2x_3$, where $x_1, x_2, x_3, x_4$ are binary variables. As to be detailed in subsequent sections, the hypergraph $G_H$ is programmable into a Rydberg atom graph using a proper set of auxiliary atoms, referred to as Rydberg hyperedges, as shown in Fig.~\ref{Fig1}(b), where the triangular subgraph between the atoms representing $x_1, x_2$, and $x_3$ corresponds to the hyperedge $(x_1, x_2, x_3)$. The HUBO solution $\mathbf{x} = (1, 0, 1, 0)$ is obtainable through a quantum adiabatic process that evolves the atom system to its many-body ground state, as depicted in Fig.~\ref{Fig1}(c).

In the rest of the paper, the new method of using Rydberg atoms to implement HUBO problems is introduced in Sec.~\ref{Sec2}. This Rydberg HUBO implementation develops two types of hyperedges: the positive-weight hyperedge and the positive-weight hyperedge, both based on the properties of Rydberg superatoms\cite{Dudin2012_superatom, Ebert2015_superatom, Zeiher2015_superatom, Labuhn2016_Ising}. In Sec.~\ref{Sec3}, applications of the Rydberg HUBO implementations are considered, including quantum simulations and quantum computing with high-order interactions. In Sec.~\ref{Sec4}, the scaling properties of the Rydberg-atom HUBO implementation are analyzed, showing that the number of atoms required is $O(N^K)$, where $N$ is the number of vertices in the hypergraph and $K$ is the maximum order of the interaction. The conclusion is presented in  Sec.~\ref{Sec5}.

\section{Higher-order Ising spin interaction} ~\label{Sec2}\noindent
We consider an extended Ising model which incorporates higher-order interactions, also known as the $p$-spin model~\cite{Derrida1980, Derrida1981}, defined as follows:
\begin{equation} \label{H}
\hat{H}=\sum_{j} J^{(1)}_j \hat{n}_j+\sum_{j<k} J^{(2)}_{jk}\hat{n}_j \hat{n}_k 
+\sum_{j<k<l } J^{(3)}_{jkl}\hat{n}_j \hat{n}_k \hat{n}_l +\cdots, 
\end{equation}
where $\hat{n}=\left|1\right>\left<1\right|$ is the number operator, taking a value of $0$ or $1$ for spin basis $\ket{0}$ or $\ket{1}$, respectively, and $J^{(j)}$ is interaction strengths where $j$-th spins involved. An Ising spin system with higher-order interactions can be naturally mapped to a hypergraph $G_H=(V,E=\{E^{(2)}, E^{(3)},\cdots \})$, where spins correspond to vertices in $V$ and $K$-th order interactions are represented by hyperedges $E^{(K)}$. Thus, the Hamiltonian $\hat{H}$ can be expressed as:
\begin{eqnarray} \label{H3}
\hat{H}&=&\sum_{j\in V} J^{(1)}_j \hat{n}_j+\sum_{(j,k)\in E^{(2)}} J^{(2)}_{jk}\hat{n}_j \hat{n}_k \nonumber \\
&&+\sum_{(j,k,l)\in E^{(3)}} J^{(3)}_{jkl}\hat{n}_j \hat{n}_k \hat{n}_l +\cdots.
\end{eqnarray}

We aim to implement $\hat{H}$ in Eq.~\eqref{H3} using a new kind of quantum wires of Rydberg atoms that effectively aggregate $K$-body interactions of Rydberg atoms. In the qubit system of Rydberg atoms, where the ground and Rydberg states are respectively represented by $\left|0\right>$ and $\left|1\right>$, the Hamiltonian governing the dynamics of a Rydberg atom graph is given by (in units of $\hbar=1$):
\begin{eqnarray} \label{HRyd}
\hat{H}_{\rm Ryd}&=&\frac{\Omega}{2} \sum_j \hat{\sigma}_j^x -\Delta \sum_j \hat{n}_j + \sum_{(j,k)} U \hat{n}_j \hat{n}_k,
\end{eqnarray}
where $\Omega$ and $\Delta$ denote the Rabi frequency and detuning of the Rydberg-atom excitation process, and the Pauli operator $\hat{\sigma}^x=\left|0\right>\left<1\right|+\left|1\right>\left<0\right|$ acts as a bit flip operator. In Hamiltonian $\hat{H}_{\rm Ryd}$,  excitation to the Rydberg state of the single atom incurs an energy penalty of $-\Delta$. When $\Omega \rightarrow 0$ and $0 < \Delta < U$, the Hamiltonian of the Rydberg atom graph becomes equivalent to the cost function of the maximum independent set (MIS) problem, which aims to maximize the occupation number ($n=1$) under the constraint of Rydberg blockade, given by $n_jn_k=0$ for $(j,k)\in E$.~\cite{Pichler2018_MIS}

The first two terms ($K=1$ and $K=2$) of $\hat{H}$ in Eq.~\eqref{H} are effectively representable in the quadratic unconstrained binary optimization (QUBO) form as:
\begin{eqnarray}
\hat{H}_{\rm QUBO} &\stackrel{g}{=}& \sum_{j \in V} J^{(1)}_j \hat{n}_j + \sum_{(j, k) \in E} J^{(2)}_{jk} \hat{n}_j \hat{n}_k,
\end{eqnarray}
where the symbol $\stackrel{g}{=}$ denotes ground state equivalence under the MIS conditions, $\Omega \rightarrow 0$ and $0 < \Delta < U$. The coefficients $J^{(1)}_{j}$ and $J^{(2)}_{jk}$ are encodable with QUBO building blocks, including the auxiliary atom set and Rydberg quantum wire~\cite{Byun2023}. In the QUBO representation, there are two kinds of quantum wires: ``even-atom quantum wire'' and ``odd-atom quantum wire'', which represent the cost function,
\begin{eqnarray}
\frac{\hat{H}^{\rm even}_{jk}}{\Delta} &\stackrel{g}{=}& \hat{n}_j \hat{n}_k \label{Eq_even} \\ 
\frac{\hat{H}^{\rm odd}_{jk}}{\Delta} &\stackrel{g}{=}& -\hat{n}_j \hat{n}_k + \hat{n}_j + \hat{n}_k \label{Eq_odd}
\end{eqnarray}

The higher-order ($K>2$) terms of $\hat{H}$ in Eq.~\eqref{H} can be viewed as hyperedges that correspond to the aggregate units of their elements, in the sense that $n_j n_k n_l \cdots = 1$ if and only if $n_j = n_k = n_l = \cdots = 1$. These higher-order terms are encodable by introducing a new kind of quantum wires for hyperedges, as to be detailed below, in such a way that a Rydberg atom graph can represent the $K$-th order hyperedge, which corresponds to the $K$-th order term in $\hat{H}$, i.e.,
\begin{eqnarray}
\hat{H}^{(K)} &\stackrel{g}{=}& \sum_{(j, k, l, \cdots) \in E^{(K)}} J_{jkl \cdots} \hat{n}_j \hat{n}_k \hat{n}_l \cdots.
\end{eqnarray}

\subsection{Higher-order unconstrained binary optimization (HUBO)}
The HUBO problem is the extended version of the QUBO problem. It includes higher-order terms in addition to the linear and quadratic terms in QUBO, targeting to obtain the solution ${\bf x} = (x_1, x_2, \cdots, x_N) \in \{0,1\}^N$ that minimizes the cost function $f({\bf x})$ defined as:
\begin{equation} \label{HUBO}
f({\bf x}) = \sum_{j} J^{(1)}_j x_j + \sum_{j<k} J^{(2)}_{jk} x_j x_k 
+ \sum_{j<k<l} J^{(3)}_{jkl} x_j x_k x_l + \cdots,
\end{equation}
where $J^{(K)}_{jkl\cdots}$ is a real-valued $K$-th order coefficient. 

\begin{figure*}[t!]
\includegraphics[width=0.85\textwidth]{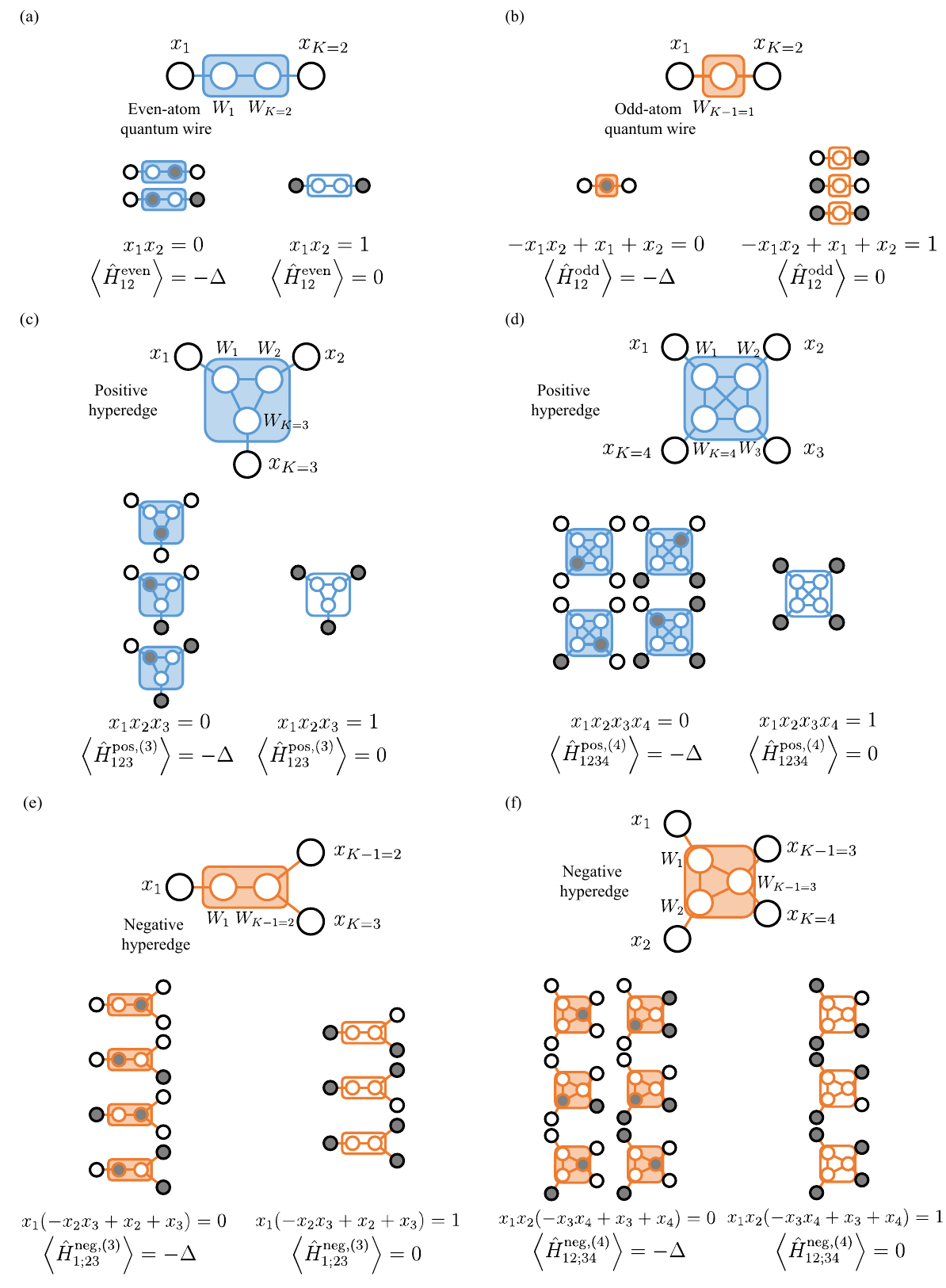}
\caption{Rydberg QUBO and HUBO implementations. (a) Positive even-atom quantum wire implementation for QUBO. (b) Negative odd-atom quantum wire implementation for QUBO. (c, d) Positive hyperedge implementations of order $K=3$ and $K=4$ using $K=3$- and $K=4$-atom Rydberg superatoms, respectively. (e, f) Negative hyperedge implementations of order $K=3$ and $K=4$ using $K-1=2$- and $K-1=3$-atom Rydberg superatoms, respectively.}
\label{Fig2}
\end{figure*}

Our approach to the implementation of $K$-th order interaction is to extend the previous QUBO implementation of Rydberg atom graphs~\cite{Byun2023}. The Rydberg QUBO implementation is illustrated in Figs.~\ref{Fig2}(a,b). The two kinds of quantum wires, the ``even-atom" quantum wire and the ``odd-atom" quantum wire, are used to encode positive and negative quadratic terms, respectively. An even-atom quantum wire configuration is shown in Fig.~\ref{Fig2}(a). This configuration connects two vertices representing variables $x_1$ and $x_2$, where $x_{1(2)} \in \{0, 1\}$, with an atom chain consisting of two atoms $W_1$ and $W_2$. In this configuration, excitation occurs in either $W_1$ or $W_2$ under the MIS condition, resulting in an additional energy of $-\Delta$ when $x_1 x_2 = 0$. Conversely, it implies an effective energy of $+\Delta$ when $x_1 x_2 = 1$. Therefore, the even-atom quantum wire has an effective energy term $\hat{H}^{\rm even}_{12}/\Delta \stackrel{g}{=} \hat{n}_1\hat{n}_2$. Likewise, Fig.~\ref{Fig2}(b) depicts the simplest odd-atom quantum wire connecting $x_1$ and $x_2$, with a single atom $W_1$. In this case, one excitation occurs in $W_1$ under the MIS condition, resulting in an effective energy of $+\Delta$ when $-x_1 x_2 + x_1 + x_2 = 1$. Therefore, the odd-atom quantum wire introduces an effective energy term $\hat{H}^{\rm odd}_{12}/\Delta \stackrel{g}{=} -\hat{n}_1\hat{n}_2+\hat{n}_1+\hat{n}_2$. 

For the implementation of higher-order terms, in the subsequent subsections, we introduce two types of hyperedges, the ``positive-weight hyperedge" and the ``negative-weight hyperedge," which correspond to the positive and negative higher-order terms, respectively. 

\subsection{Positive-weight hyperedge}
A $K$-th order positive hyperedge is a ``positive''-weighted aggregation of $K$ vertices, meaning $J_{123\cdots} > 0$ for $\left(x_1,x_2,x_3, \cdots \right) \in E^{(K)}$. Similar to the even-atom quantum wire in QUBO, making a subgraph which satisfies 
\begin{eqnarray} \label{Eq_pos}
\frac{\hat{H}^{\rm pos}_{123\cdots}}{\Delta} \stackrel{g}{=} \prod_{j=1}^{K} \hat{n}_j,
\end{eqnarray}
where the set of auxiliary atoms act as a positive-weight hyperedge. This condition can be generated by the $K$-atom Rydberg superatom~\cite{Dudin2012_superatom, Ebert2015_superatom, Zeiher2015_superatom, Labuhn2016_Ising}, a cluster of atoms that share the Rydberg blockade regime. By the character of the Rydberg blockade, a Rydberg superatom only permits single-atom excitation under the MIS condition.

Figure~\ref{Fig2}(c) showcases a positive-weight hyperedge with the maximum order of $K=3$. The $K=3$-atom Rydberg superatom, forming a triangle with atoms $W_1$, $W_2$, and $W_3$, which is connected to the three vertices $x_1$, $x_2$, and $x_3$, acting as a hyperedge $(x_1,x_2,x_3)$ with a positive energy contribution. Under the MIS condition, only a single excitation in the $K=3$ Rydberg superatom is permitted, adding an energy penalty of $-\Delta$, where $x_1x_2x_3=0$. Conversely, no excitation occurs in the Rydberg superatom when $x_1x_2x_3=1$, thereby assigninig a positive energy of $+\Delta$ to the spin configuration corresponding to $x_1x_2x_3=1$. Thus, the energy function incorporating the additional cost from the Rydberg superatom is represented as $\hat{H}^{\rm pos, (3)}_{123} /\Delta = \hat{n}_1 \hat{n}_2 \hat{n}_3$, following the form in Eq.~\eqref{Eq_pos}. 

Similarly, in Fig.~\ref{Fig2}(d), the hyperedge with the maximum order of $K=4$ is illustrated. The $K=4$ Rydberg superatom in Fig.~\ref{Fig2}(d) generates $\hat{H}^{\rm pos, (4)}_{1234} /\Delta = \hat{n}_1 \hat{n}_2 \hat{n}_3 \hat{n}_4$ term.
The many-body ground states of the connected Rydberg superatom under the MIS condition are listed in Table~\ref{Tab_pos}.

\begin{table}
\centering
\caption{The many-body ground state energies of positive-weight hyperedges, generated by $K$-atom Rydberg superatoms under the MIS condition, depend on the spin configuration of data qubits $x_1, x_2, \cdots, x_K$.}
\begin{ruledtabular}
\begin{tabular}{@{}ccc@{}}
$\left|x_1 x_2 \cdots x_K\right>$ & Hyperedge configuration & Energy  \\
\hline
$\left|00\cdots 0\right>$ & $(\left|00\cdots 1\right>+\cdots +\left|10\cdots 0\right>)/\sqrt{K}$ & $-\Delta$ \\
$\left|00\cdots 1\right>$ &  $(\left|0\cdots 10\right>+\cdots +\left|10\cdots 0\right>)/\sqrt{K-1}$ & $-\Delta$ \\
$\vdots$ & $\vdots$ & $\vdots$ \\
$\left|11\cdots 0\right>$ & $\left| 00\cdots 1\right>$ & $-\Delta$ \\
$\left|11\cdots 1\right>$ & $\left| 00\cdots 0\right>$ & $0$ \\
\end{tabular}
\end{ruledtabular}
\label{Tab_pos}
\end{table}

It is noted that the even-atom quantum wire, one of the QUBO building blocks, also follows the positive-weight hyperedge implementation. QUBO is the case containing $K = 2$ order terms as the maximum order term. The $K = 2$-atom Rydberg superatom is a dimer, one of the simplest even-atom chains, as shown in Fig.~\ref{Fig2}(a). The energy corresponding to the even-atom quantum wire in Eq.~\eqref{Eq_even} also satisfies Eq.~\eqref{Eq_pos}, i.e., $\hat{H}^{\rm even}_{jk}=\hat{H}^{{\rm pos}, (K=2)}_{jk}$.

\subsection{Negative-weight hyperedge}
A $K$-th order negative-weight hyperedge is a ``negative"-weighted aggregation of $K$ vertices, meaning $J_{123\cdots} < 0$ for $\left(x_1, x_2, x_3, \cdots \right) \in E^{(K)}$. To achieve this, the term $-\prod_{j=1}^{K} \hat{n}_j$ should be implemented. Consequently, the effective Hamiltonian of the new quantum wire must incorporate $-\prod_{j=1}^{K} \hat{n}_j$. The $(K-1)$-atom Rydberg superatom, which is connected with $K$ vertices, possesses such an effective Hamiltonian. So, we connect all elements of the Rydberg superatom, from $W_1$ to $W_{K-1}$, with vertices $x_1$ through $x_{K-1}$, and also connect the final vertex $x_K$ to $W_{K-1}$, which is shared with the $x_{K-1}$-th vertex, as in Figs.~\ref{Fig2}(e,f). Then, under the MIS condition, to prevent excitation of the atoms in the Rydberg superatom, vertices from $x_1$ to $x_{K-2}$ must be excited, and vertices $K-1$ and $K$ must satisfy $-x_{K-1}x_{K} + x_{K-1} + x_{K} = 1$. If this condition is not met, there will be a single-atom excitation in the Rydberg superatom, effectively contributing an energy of $-\Delta$. Thus, when the conditions $x_1x_2\cdots x_{K-2}=1$ and $-x_{K-1}x_{K} + x_{K-1} + x_{K} = 1$ are met, specifically, for $(x_{K-1}, x_K) = (0,1)$, $(x_{K-1}, x_K) = (1,0)$, and $(x_{K-1}, x_K) = (1,1)$, the Rydberg superatom has an effective energy of $+\Delta$. The corresponding Hamiltonian is given by
\begin{equation}
\frac{\hat{H}^{\rm neg}_{12\cdots(K-2);(K-1)K}}{\Delta} \stackrel{g}{=} \left[-\hat{n}_{K-1}\hat{n}_K + \hat{n}_{K-1} + \hat{n}_K\right] \prod_{j=1}^{K-2} \hat{n}_j,
\label{Eq_neg}
\end{equation}
which contains the negative $K$-th order term and two additional positive $(K-1)$-th order terms.

Figure~\ref{Fig2}(e) illustrates a negative-weight hyperedge with the maximum order of $K=3$. The $K-1=2$-atom Rydberg superatom, forming a dimer with atoms $W_1$ and $W_2$ is connected to the vertices $x_1$, $x_2$, and $x_3$, acting as a hyperedge $(x_1, x_2, x_3)$ with a negative energy contribution. Under the MIS condition, only a single excitation in the $K-1=2$ Rydberg superatom is allowed, adding an energy penalty of $-\Delta$, when $x_1(-x_2x_3 + x_2 + x_3) = 0$. Conversely, no excitation occurs in the Rydberg superatom when $x_1(-x_2x_3 + x_2 + x_3) = 1$, thereby assigning a positive energy of $+\Delta$ to the spin configuration corresponding to $x_1(-x_2x_3 + x_2 + x_3) = 1$. Thus, the energy function incorporating the additional cost from the Rydberg superatom is given by $\hat{H}^{\rm neg, (3)}_{1;23} /\Delta = \hat{n}_1 (-\hat{n}_2 \hat{n}_3 + \hat{n}_2 + \hat{n}_3)$, following the form in Eq.~\eqref{Eq_neg}. 

Similarly, the $K-1=3$ Rydberg superatom in Fig.~\ref{Fig2}(f) generates $\hat{H}^{\rm neg, (4)}_{12;34} /\Delta = \hat{n}_1 \hat{n}_2 (-\hat{n}_3 \hat{n}_4 + \hat{n}_3 + \hat{n}_4)$ term. The many-body ground states of the connected Rydberg superatom under the MIS condition are listed in Table~\ref{Tab_neg}.

\begin{table}
\centering
\caption{The many-body ground state energies of negative-weight hyperedges, generated by $(K-1)$-atom Rydberg superatoms under the MIS condition, depend on the spin configuration of data qubits $x_1, x_2, \cdots, x_K$.}
\begin{ruledtabular}
\begin{tabular}{@{}cccc@{}}
$\left|x_1  \cdots x_{K-2};x_{K-1}x_K\right>$ & Hyperedge configuration  && Energy  \\
\hline
$\left|0\cdots 0;00\right>$ & \multicolumn{2}{l}{\begin{tabular}{ll} $(\left|0\cdots 001\right>+\cdots $ \\ $ +\left|1\cdots 000\right>)/\sqrt{K-1}$ \end{tabular}} 
& $-\Delta$ \\
$\left|0\cdots 0;01\right>$ & \multicolumn{2}{l}{\begin{tabular}{ll} $(\left|0\cdots 010\right>+\cdots $ \\ $+\left|1\cdots 000\right>)/\sqrt{K-2}$ \end{tabular}}  
& $-\Delta$ \\
$\vdots$ & $\vdots$ && $\vdots$ \\
$\left|1\cdots 1;00\right>$ & $\left| 0\cdots 001\right>$ && $-\Delta$ \\
$\left|1\cdots 1;01\right>$ & $\left| 0\cdots 000\right>$ && $0$ \\
$\left|1\cdots 1;10\right>$ & $\left| 0\cdots 000\right>$ && $0$ \\
$\left|1\cdots 1;11\right>$ & $\left| 0\cdots 000\right>$ && $0$ \\
\end{tabular}
\end{ruledtabular}
\label{Tab_neg}
\end{table}

It is noted that if $K=2$, the $\prod_{j=1}^{K-2} \hat{n}_j$ term in Eq.~\eqref{Eq_neg} can be omitted, leaving only the terms $-\hat{n}_1\hat{n}_2 + \hat{n}_1 + \hat{n}_2$, which match Eq.~\eqref{Eq_odd}. This indicates that the negative-weight hyperedge in Rydberg HUBO implementation includes the odd-atom quantum wire in Rydberg QUBO implementation, such that $\hat{H}^{{\rm odd}}_{jk}=\hat{H}^{{\rm neg},(K=2)}_{jk}$.

\section{Programming Rydberg atom graphs for HUBO problems} \label{Sec3}

\begin{figure*}[hbt!]
\includegraphics[width=2.0\columnwidth]{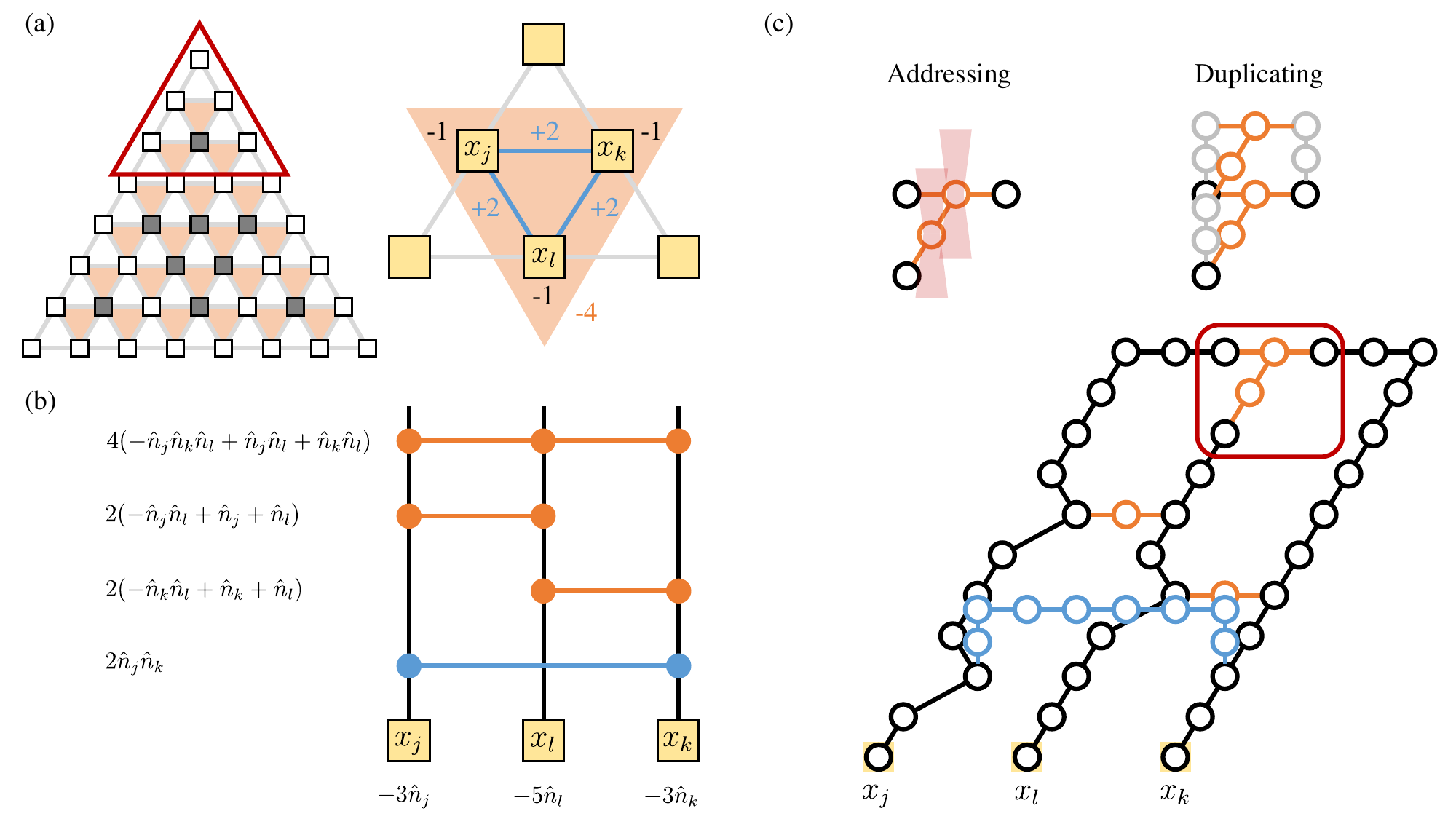}
\caption{(a) Sierpinski triangle shaped spin ordering. The downward-facing triangle is subject to a third-order interaction $\hat{H}_{\rm ST}=-J \hat{\sigma}_j^z\hat{\sigma}_k^z\hat{\sigma}_l^z$. The corresponding hypergraph represents the HUBO cost function.
(b) A sketch of the Rydberg atom graph corresponding to the downward-facing triangle. It includes three quantum wires and one third-order hyperedge. (c) The skeleton Rydberg atom graph, which is the unweighted version of the graph.}
\label{Fig3}
\end{figure*}

HUBO problems can be transformed into QUBO problems~\cite{Zaman_IEEE_2022, Mandal_2020, DwaveHandbook} without utilizing the hypergraph implementation. However, converting the higher-order terms in a HUBO problem into quadratic terms for a QUBO problem necessitates additional variables, thereby increasing the number of required atoms. While this increase can be polynomially bounded in specific cases~\cite{Boros_DAM_2002}, the number of auxiliary variables generally grows exponentially, significantly increasing the resources needed to solve the HUBO problem~\cite{Rodrigueze-Heck_thesis_2018}. This resource increase must be considered when transforming HUBO problems to QUBO, as exponential growth in resources can make the problem significantly more challenging to solve. A direct HUBO implementation is thus crucial to avoid the additional atom resources required by the transformation from HUBO to QUBO.

Now, the HUBO problem can be encoded into a Rydberg atom graph by utilizing the positive-weight hyperedge defined in Eq.\eqref{Eq_pos} and the negative-weight hyperedge defined in Eq.\eqref{Eq_neg}, with appropriately tuned weights:
\begin{align} \label{eq_HUBO_term}
&\hat{H}_{\rm HUBO} = 
\sum_{(j)} w^{\rm data}_j \hat{H}^{\rm data}_j + w^{\rm offset}_j \hat{H}^{\rm offset}_j   \nonumber \\ 
&+ \sum_{E^{(K)}, K} w^{{\rm pos}, (K)}_{(j,k,l,\cdots)} \hat{H}^{{\rm pos}, (K)}_{(j,k,l,\cdots)} + w^{{\rm neg}, (K)}_{(j,k,l,\cdots)} \hat{H}^{{\rm neg}, (K)}_{(j,k,l,\cdots)},
\end{align}
where $\hat{H}^{\rm data}$ and $\hat{H}^{\rm offset}$ are Hamiltonians corresponding to data and offset qubits, respectively, which are components of the QUBO building blocks~\cite{Byun2023} encoding linear terms. The weights $w$ determine the coupling strength of each term and can be set using local laser beam addressing~\cite{Labuhn2014_addressing, Omran2019_20addressing, Graham2022_MAXCUT, deOliveira2024_MWIS} or through duplication~\cite{Byun2023} with 3D stacking~\cite{Lee2016_3Drearrange, Barredo2018_3D, Kim2020_3DRyd, Song2021_Cayleytree}. To implement HUBO with locally focused light, a weighted detuned beam should be applied to all the atoms in the Rydberg superatom, which serves as the hyperedge gadget.

In the following, we consider two experimentally feasible candidates that necessitate higher-order interactions. The first involves the quantum simulation of complex spin systems, and the second relates to the application of HUBO-based adiabatic quantum computing.

\subsection{Quantum Sierpinski triangle}
When the downward-facing triangles in a triangular lattice follow the Hamiltonian  $\hat{H}_{\rm ST}$,
\begin{eqnarray}
\hat{H}_{\rm ST}=-J\sum_{\triangledown_{jkl}}\hat{\sigma}^z_{j} \hat{\sigma}^z_{k} \hat{\sigma}^z_{l},
\end{eqnarray}
the ground state contains an odd number of up-spins $\left|1\right>$ in each downward-facing triangle, where $\sigma_z$ is the pauli $z$ operator. This configuration satisfies  $\hat{\sigma}^z_{j} \hat{\sigma}^z_{k} \hat{\sigma}^z_{l}=+1$ under the ferromagnetic condition $J>0$, where $\hat{\sigma}_z=-\left|0\right>\left<0\right|+\left|1\right>\left<1\right|$ is the Pauli operator. The many-body ground state of the  spin system forms the shape of a Sierpinski triangle, shown in Fig.~\ref{Fig3}(a), which is a characteristic fractal structure~\cite{NEMJ2022_fractal, Newman1999_ST}. Then, the Hamiltonian $\hat{H}_{\rm ST}$ can be expressed using a Rydberg atom graph as follows:
\begin{eqnarray}\label{Sierpinski_HUBO}
\hat{H}_{\rm ST} \propto \sum_{\triangledown_{jkl}} & &\left[-4\hat{n}_j\hat{n}_k\hat{n}_l + 2\hat{n}_j\hat{n}_k + 2\hat{n}_k\hat{n}_l + 2\hat{n}_j\hat{n}_l \right. \nonumber \\ 
& & \left. -\hat{n}_j - \hat{n}_k - \hat{n}_l \right].
\label{sierpinski}
\end{eqnarray}

Figure~\ref{Fig3}(b) depicts the illustration of the Rydberg atom graph~\cite{Qiu2020} representing a unit downward-facing triangle, which is highlighted in Fig.~\ref{Fig3}(a). The black lines represent antiferro (AF)-ordered quantum wires~\cite{Kim2022_wire, Byun2022PRXQ_PlatonicSolid, Byun2023} that facilitate the establishment of non-local interactions. The Hamiltonian for the unit downward-facing triangle in Eq.~\eqref{sierpinski} can be formulated in the format of Eq.~\eqref{eq_HUBO_term} as
\begin{eqnarray}
\hat{H}^{\rm ST}_{\triangledown_{jkl}} & = & w^{\rm data}_j \hat{H}^{\rm data}_j + w^{\rm data}_k \hat{H}^{\rm data}_k + w^{\rm data}_l \hat{H}^{\rm data}_l \nonumber \\
 &+& w^{\rm neg,(2)}_{jl} \hat{H}^{\rm neg,(2)}_{jl} + w^{\rm neg,(2)}_{kl} \hat{H}^{\rm neg,(2)}_{kl} \nonumber \\
 &+& w^{\rm pos,(2)}_{jk} \hat{H}^{\rm pos,(2)}_{jk} + w^{\rm neg,(3)}_{jk;l} \hat{H}^{\rm neg,(3)}_{jk;l},
\end{eqnarray}
where the weights are $w^{\rm data}j = w^{\rm data}_{k = 3}$, $w^{\rm data}_{l = 5}$, $w^{\rm pos,(2)}_{jk} = w^{\rm neg,(2)}_{jl} = w^{\rm neg,(2)}_{kl} = 2$, and $w^{\rm neg,(3)}_{jkl} = 4$. In Fig.~\ref{Fig3}(c), a skeleton Rydberg atom graph is depicted where all weights are $w = 1$, corresponding to the sketch in Fig.~\ref{Fig3}(b). The highlighted region in Fig.~\ref{Fig3}(c) contains the third-order negative hyperedge, utilizing the same configuration as in  Fig.~\ref{Fig2}(e). As illustrated in Fig.~\ref{Fig3}(c), weights can be assigned through local beam addressing or by duplicating the hyperedge subgraph.

\begin{figure*}[bht!]
\includegraphics[width=2.0\columnwidth]{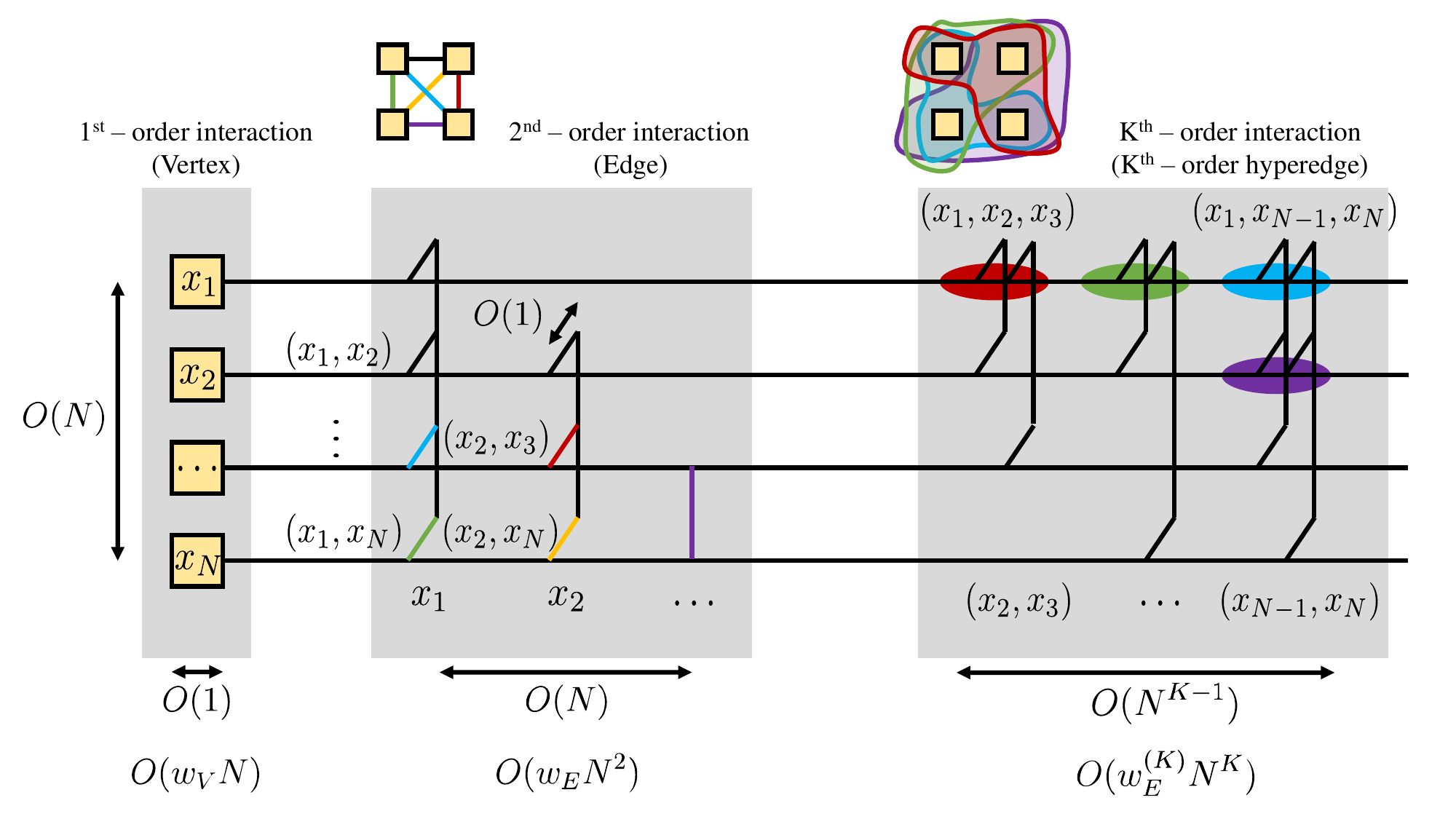}
\caption{The diagram depicts a fully connected hypergraph. In QUBO cases, edges are represented by quantum wires, shown as colored lines in the figure. In HUBO cases, hyperedges are represented by the hyperedge gadgets introduced in this paper, illustrated as colored circles in the figure. To represent a $K$-th order hypergraph, the duplication method requires $O(\sum_{K=1} w_E^{(K)} N^K)$ atoms.}
\label{Fig4}
\end{figure*}

\subsection{Factorization problems}
Using HUBO enables the solution of the factorization problem. The objective of prime factorization is to identify integers $P$ and $Q$ for a given integer $N$, such that $N = P \times Q$. For example, to factor $N = 6 = (110)_2$, where $(N_{m} \cdots N_1 N_0)_2$ denotes the binary notation and $N = N_{m} 2^{m} + \cdots + N_1 2^1 + N_0 2^0$, the cost function of the factorization problem is expressed as
\begin{eqnarray}
f_{\rm Fact6} (\mathbf{x}) &=& \left[6 - (2^1 P_1 + 2^0 P_0)(2^1 Q_1 + 2^0 Q_0) \right]^2.
\label{factorization}
\end{eqnarray}
To simplify the problem, let $P_0 = 1$. Subsequently, the cost function transforms to:
\begin{eqnarray}
f_{\rm Fact6} (\mathbf{x}) &=& -20 Q_1 - 11 Q_0 - 16 P_1 Q_1 - 16 P_1 Q_0 \nonumber \\ 
&& + 4 Q_1 Q_0 + 32 P_1 Q_1 Q_0.
\label{factorization_simplified}
\end{eqnarray}
This constitutes a three-variable optimization problem involving $P_1$, $Q_1$, and $Q_0$. Similar to the previous example, we express the Hamiltonian in the form of Eq.~\eqref{eq_HUBO_term}:
\begin{eqnarray}
\hat{H}_{\rm Fact 6} & = & w^{\rm data}_{P_1} \hat{H}^{\rm data}_{P_1} + w^{\rm data}_{Q_1} \hat{H}^{\rm data}_{Q_1} + w^{\rm data}_{Q_0} \hat{H}^{\rm data}_{Q_0} \nonumber \\
 &+& w^{\rm neg,(2)}_{P_1Q_1} \hat{H}^{\rm neg,(2)}_{P_1Q_1} + w^{\rm neg,(2)}_{P_1Q_0} \hat{H}^{\rm neg,(2)}_{P_1Q_0} \nonumber \\
 &+& w^{\rm pos,(2)}_{Q_1Q_0} \hat{H}^{\rm pos,(2)}_{Q_1Q_0} + w^{\rm pos,(3)}_{P_1Q_1Q_0} \hat{H}^{\rm pos,(3)}_{P_1Q_1Q_0},
\end{eqnarray}
which involves four different quantum wires and hyperedges: a $K=3$ order positive hyperedge for $32 P_1 Q_1 Q_0$, odd-atom quantum wires for $16(-P_1 Q_1 + P_1 + Q_1)$ and $16(-P_1 Q_0 + P_1 + Q_0)$, and an even-atom quantum wire for $4 Q_1 Q_0$. The weight factors are $w^{\rm data}_{P_1}=32$, $w^{\rm data}_{Q_1}=36$, $w^{\rm data}_{Q_0}=27$, $w^{\rm neg,(2)}_{P_1Q_1} = w^{\rm neg,(2)}_{P_1Q_0}=16$, $w^{\rm pos,(2)}_{Q_1Q_0}=4$, and $w^{\rm pos,(3)}_{P_1Q_1Q_0}=32$. The solution to the HUBO is $(P_1; Q_1, Q_0) = (1; 1, 0)$, corresponding to $P = (11)_2 = 3$ and $Q = (10)_2 = 2$, satisfying $6 = 2 \times 3$. This solution can be obtained via quantum adiabatic passage to the MIS condition.

\section{Discussion}~\label{Sec4}
To discuss the scaling behavior, we employ a 3D quantum wire lattice structure~\cite{Qiu2020}, akin to Figs.~\ref{Fig3}(b)-(c). A line of $N$ data qubits is organized, with quantum wires effectively duplicating this line of data qubits. Hence, the scalability is determined by the number of duplications.

In Fig.~\ref{Fig4}, the Rydberg atom graph is illustrated, where black lines represent antiferro (AF)-ordered quantum wires~\cite{Kim2022_wire, Byun2022PRXQ_PlatonicSolid, Byun2023}, connecting atoms such that vertices effectively link together. Branches extend from AF-ordered quantum wires, running parallel to the vertex line and crossing over other AF-ordered quantum wires to form edges of the graph. In Fig.~\ref{Fig4}, a branch originates from $x_1$, passes over $x_2$, $\cdots$, $x_N$, and connects to all others, ensuring $x_1$-to-all  connectivity. Similarly, branches are extended from $x_2$ to $x_{N-1}$, and so forth, achieving an all-to-all connected graph. The number of branches scales as $O(N)$ while the height scales as $O(1)$~\cite{Qiu2020}. Therefore, to generate an all-to-all connected unweighted graph, $O(N^2)$ atoms are required\cite{Qiu2020}. In the case of a weighted graph, local addressing and duplication methods necessitate $O(N^2)$ atoms and $O(w_V N+w_E N^2)$, respectively, where $w_V$ and $w_E$ denote the maximum weight of vertices and edges~\cite{Qiu2020, Byun2023}.

To implement hyperedges of the highest order $K$ on a cubic lattice structure, branches are utilized to represent combinations of vertices. For instance, Fig.~\ref{Fig4} illustrates a hypergraph with the maximum degree $K=3$, where the colored circles indicate hyperedges. To implement a $K=3$ order hyperedge, branches are formed by selecting $K-1$ vertices from $N-1$. The number of branches scales as $O(N^{K-1})$ scaling. Ultimately, constructing a hypergraph of order $K$ requires $O(N^K)$ atoms. Similar to QUBO, in the case of implementing a weighted graph, the required number of atoms is $O\left(\sum_{K=1} w_E^{(K)} N^K\right)$, where $w_E^{(K)}$ denotes the maximum weight of the $K$-th hyperedges, with $w_E^{(K=1)}=w_V$ and $w_E^{(K=2)}=w_E$.

\begin{figure}[t]
\includegraphics[width=0.5\textwidth]{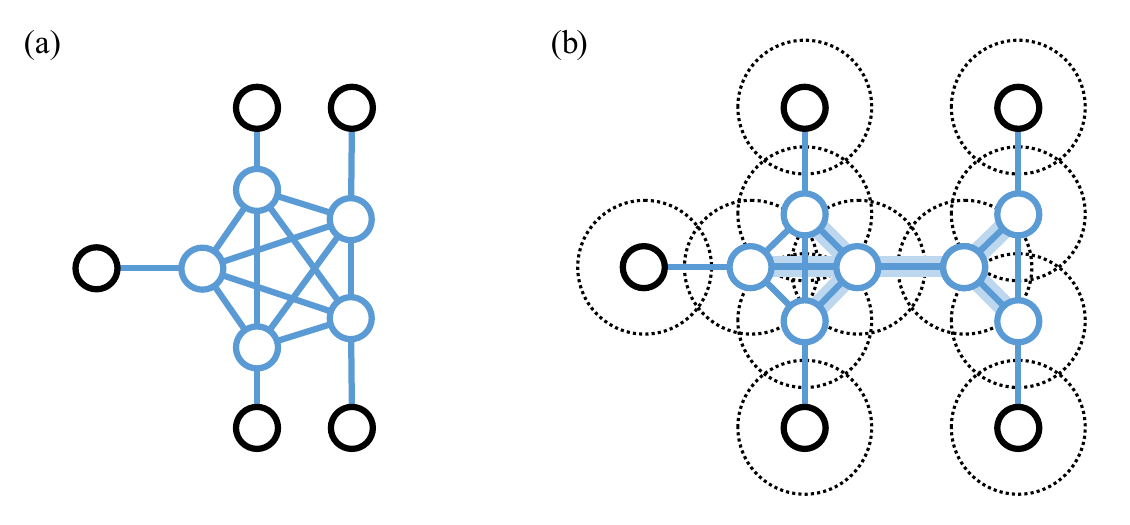}
\caption{Hyperedges implemented using AF-ordered quantum wires. (a) The target graph with a $K=5$ positive-weight hyperedge. (b) By employing AF-ordered quantum wires and vertex splitting, a Rydberg superatom is realized using  $O(K^2)$ atoms.}
\label{Fig5}
\end{figure}

If $K$, the order of the interaction, becomes larger, implementing a Rydberg superatom becomes increasingly challenging. However, using an AF-ordered quantum wire and the vertex splitting method~\cite{Kim2022_wire, Byun2022PRXQ_PlatonicSolid, Byun2023}, it is possible to implement a superatom-equivalent graph. Figure \ref{Fig5}(a) shows the target graph corresponding to the positive-weight hyperedge with $K=5$. Figure \ref{Fig5}(b) displays the Rydberg atom graph used for hyperedge implementation. A Rydberg superatom is a type of Rydberg atom graph, which can be programmed using Rydberg atom graph QUBO implementation. For a $K$-atom Rydberg atom graph, the number of vertices is $K$, so each hyperedge requires $O(K^2)$ atoms. The number of hyperedges in a $K$-th order hypergraph is $O(N^K)$, thus requiring $O(K^2 N^K)$ atoms. If $K$ is a finite number, the scaling remains $O(N^K)$.

\section{Conclusion}~\label{Sec5}
Rydberg-atom graph gadgets are introduced to efficiently program $K$-th order interactions within a Rydberg atom system under the MIS condition. This methodology facilitates the determination of many-body ground states for Ising-type Hamiltonians, which are encoded to tackle HUBO, the higher-order unconstrained optimization problem. This Rydberg-atom approach extends beyond solving classical optimization problems to quantum simulations of spin models. The polynomial scaling of $O(N^K)$, in terms of the number of atoms required for $N$-vertex hypergraph optimization problems underscores the experimental feasibility of Rydberg atom-based higher-order graph optimization using current and near-term devices.

\begin{acknowledgements}
We thank Jinhyung Lee for fruitful discussions.
\end{acknowledgements}


\begin{thebibliography}{1}



\bibitem{Efimov1970} V. Efimov,
	``Energy levels arising from resonant two-body forces in a three-body system,''
	Phys. Lett. B {\bf33}, 563-564 (1970).

\bibitem{Naidon2017_Efimov} P. Naidon, and S. Endo,
	``Efimov physics: a review,''
	Rep. Prog. Phys. {\bf80}, 056001 (2017).

\bibitem{Cooper2004_FQH} N. R. Cooper,
	``Exact Ground States of Rotating Bose Gases Close to a Feshbach Resonance,''
	Phys. Rev. Lett. {\bf 92}, 220405 (2004).

\bibitem{Levin2005_stringnet} M. A. Levin and X.-G. Wen,
		``String-net condensation: A physical mechanism for topological phases,''
		Phys. Rev. B {\bf 71}, 045110 (2005).

\bibitem{Levin2005_stringnet2} M. A. Levin and X.-G. Wen,
		``Colloquium: Photons and electrons as emergent phenomena,''
		Rev. Mod. Phys. {\bf 77}, 871 (2005).


\bibitem{Buchler2007_3polar} H. P. B{\"u}chler, A. Micheli and P. Zoller,
		``Three-body interactions with cold polar molecules," 
		Nat. Phys {\bf3}, 726-731 (2007).

\bibitem{Schmidt2008_3B2Dlattice} K. P. Schmidt, J. Dorier, and A. M. L{\"a}uchli,
		``Solids and Supersolids of Three-Body Interacting Polar Molecules on an Optical Lattice," 
		Phys. Rev. Lett. {\bf101}, 150405 (2008).

\bibitem{BCS2009_3B1Dlattice} B. Capogrosso-Sansone, S. Wessel, H. P. B{\"u}chler, P. Zoller, and G. Pupillo,
		``Phase diagram of one-dimensional hard-core bosons with three-body interactions," 
		Phys. Rev. B. {\bf79}, 020503(R) (2009).

\bibitem{Bonnes2009_3B1Dlattice} L. Bonnes, H. B{\"u}chler, and S. Wessel,
		``Polar molecules with three-body interactions on the honeycomb lattice," 
		New J. Phys. {\bf12}, 053027 (2010).









		
\bibitem{Seeley2012_H2} J. T. Seeley, M. J. Richard, and P. J. Love,
``The Bravyi-Kitaev transformation for quantum computation of electronic structure,''
J. Chem. Phys. {\bf137}, 224109 (2012)

\bibitem{Hauke2013_Schwinger} P. Hauke, D. Marcos, M. Dalmonte, and P. Zoller,
	``Quantum Simulation of a Lattice Schwinger Model in a Chain of Trapped Ions,''
	Phys. Rev. X {\bf3}, 041018 (2013).

\bibitem{Pedersen2021_LGT} S. P. Pedersen and N. T. Zinner
	``Lattice gauge theory and dynamical quantum phase transitions using noisy intermediate-scale quantum devices,''
	Phys. Rev. B {\bf103}, 235103 (2021).

\bibitem{Farrell2023_QCD} R. C. Farrell, I. A. Chernyshev, S. J. M. Powell, N. A. Zemlevskiy, M. Illa, and M. J. Savage,
``Preparations for quantum simulations of quantum chromodynamics in 1+1 dimensions. I. Axial gauge,''
Phys. Rev. D {\bf107}, 054512 (2023).


\bibitem{Rossi2013_hyper} M. Rossi, M. Huber, D. Bru{\ss}, and C. Macchiavello,
		Quantum hypergraph states,
		New J. Phys. {\bf15}, 113022 (2013).

\bibitem{Liu2022_hyper} Z.-W. Liu and A. Winter, 
		Many-Body Quantum Magic,
		PRX Quantum {\bf3}, 020333 (2022).

\bibitem{Kitaev2003} A. Y. Kitaev,
	``Fault-tolerant quantum computation by anyons.''
	Ann. Phys. {\bf303}, 2-30 (2003).
\bibitem{Paetznick2013} A. Paetznick and B. W. Reichardt,
``Universal Fault-Tolerant Quantum Computation with Only Transversal Gates and Error Correction,''
Phys. Rev. Lett. {\bf111}, 090505 (2013).

\bibitem{Bluvstein2022_Toric} D. Bluvstein, H. Levine, G. Semeghini, T. T. Wang, S. Ebadi, M. Kalinowski, A. Keesling, N. Maskara, H. Pichler, M. Greiner, V. Vuleti{\'c}, and M. D. Lukin, 
		``A quantum processor based on coherent transport of entangled atom arrays,''
		Nature {\bf604}, 451–456 (2022).

\bibitem{Google2023} Google Quantum AI,
		``Suppressing quantum errors by scaling a surface code logical qubit,''
		Nature {\bf614}, 676-681 (2023).

\bibitem{Bluvstein2024_logical} D. Bluvstein, S. J. Evered, A. A. Geim, S. H. Li, H. Zhou, T. Manovitz, S. Ebadi, M. Cain, M. Kalinowski, D. Hangleiter, J. P. B. Ataides, N. Maskara, I. Cong, X. Gao, P. S. Rodriguez, T. Karolyshyn, G. Semeghini, M. J. Gullans, M. Greiner, V. Vuleti{\'c}, and M. D. Lukin,
		``Logical quantum processor based on reconfigurable atom arrays,''
		Nature {\bf626}, 58-65 (2024).

\bibitem{Self2024} C. N. Self, M. Benedetti, and D. Amaro
	``Protecting expressive circuits with a quantum error detection code,''
	Nat. Phys {\bf20}, 219-224 (2024).

\bibitem{Iqbal2024} M. Iqbal, N. Tantivasadakarn, R. Verresen, S. L. Campbell, J. M. Dreiling, C. Figgatt, J. P. Gaebler, J. Johansen, M. Mills, S. A. Moses, J. M. Pino, A. Ransford, M. Rowe, P. Siegfried, R. P. Stutz, M. Foss-Feig, A. Vishwanath and H. Dreyer
	``Non-Abelian topological order and anyons on a trapped-ion processor,''
	Nature {\bf626}, 505-511 (2024).


\bibitem{Browaeys2020} A. Browaeys and T. Lahaye,
		“Many-body physics with individually controlled Rydberg atoms,” 
		Nat. Phys. {\bf16}, 132-142 (2020).






\bibitem{Jaksch2000_blockade}  D. Jaksch, J. I. Cirac, P. Zoller, S. L. Rolston, R. C{\^o}t{\'e}, and M. D. Lukin, 
	      ``Fast Quantum Gates for Neutral Atoms'', 
	       Phys. Rev. Lett. {\bf85}, 2208 (2000).

\bibitem{Lukin2001_blockade} M. D. Lukin, M. Fleischhauer, R. Cote, L. M. Duan, D. Jaksch, J. I. Cirac and P. Zoller,
		``Dipole Blockade and Quantum Information Processing in Mesoscopic Atomic Ensembles,''
		Phys. Rev. Lett. {\bf87}, 037901 (2001).

\bibitem{UrbanNatPhys2009_blockade} E. Urban, T. A. Johnson, T. Henage, L. Isenhower, D. D. Yavuz, T. G. Walker and M. Saffman, 
	      ``Observation of Rydberg blockade between two atoms,'' 
	       Nat. Phys. {\bf5}, 110-114 (2009).

\bibitem{GaetanNatPhys2009_blockade} A. Ga\"{e}tan, Y. Miroshnychenko, T. Wilk, A. Chotia, M. Viteau, D. Comparat, P. Pillet, A. Browaeys and P. Grangier, 
		``Observation of Collective Excitation of Two Individual Atoms in the Rydberg Blockade Regime,'' 
		Nat. Phys. {\bf5}, 115-118 (2009).

\bibitem{Glaetzle2017_LHZ} A.W. Glaetzle, R. M. W. van Bijnen, P. Zoller and W. Lechner, 
		``A coherent quantum annealer with Rydberg atoms,'' 
		Nat. Commun. {\bf 8}, 15813 (2017).

\bibitem{Gambetta2020_3Ryd} F. M. Gambetta, W. Li, F. Schmidt-Kaler, and I. Lesanovsky,
		``Engineering NonBinary Rydberg Interactions via Phonons in an Optical Lattice," 
		Phys. Rev. Lett. {\bf124}, 043402 (2020).

\bibitem{Pohl2009_AB} T. Pohl, and P. R. Berman,
	``Breaking the dipole blockade: nearly resonant dipole interactions in few-atom systems,''
	Phys. Rev. Lett. {\bf102}, 013004 (2009).

\bibitem{Faoro2015_Forster} R. Faoro, B. Pelle, A. Zuliani, P. Cheinet, E. Arimondo, and P. Pillet
	``Borromean three-body FRET in frozen Rydberg gases,''
	Nat. Commun. {\bf6}, 8173 (2015).

\bibitem{Ryabtsev2018_Forster} I. I. Ryabtsev, I. I. Beterov, D. B. Tretyakov, E. A. Yakshina, V. M. Entin, P. Cheinet, and P. Pillet
	``Coherence of three-body Förster resonances in Rydberg atoms,''
	Phys. Rev. A {\bf98}, 052703 (2018).

\bibitem{Gurian2012_Forster} J. H. Gurian, P. Cheinet, P. Huillery, A. Fioretti, J. Zhao, P. L. Gould, D. Comparat, and P. Pillet,
	``Observation of a Resonant Four-Body Interaction in Cold Cesium Rydberg Atoms,''
	Phys. Rev. Lett {\bf108}, 023005 (2012).

\bibitem{NEMJ2022_fractal} N. E. Myerson-Jain, S. Yan , D. Weld, and C. Xu
``Construction of Fractal Order and Phase Transition with Rydberg Atoms,''
Phys. Rev. Lett {\bf128}, 017601 (2022).




\bibitem{Kim2022_wire} M. Kim, K. Kim, J. Hwang, E.-G. Moon, and J. Ahn, 
		``Rydberg quantum wires for maximum independent set problems,''  
		Nat. Phys {\bf18}, 755-759 (2022).

\bibitem{Byun2022PRXQ_PlatonicSolid} A. Byun, M. Kim, and J. Ahn,
             ``Finding the maximum independent sets of Platonic graphs using Rydberg atoms,''
             PRX Quantum {\bf3}, 030305 (2022).

\bibitem{Qiu2020} X. Qiu, P. Zoller, and X. Li , ``Programmable Quantum Annealing Architectures with Ising Quantum Wires,'' PRX Quantum {\bf1}, 020311 (2020)


\bibitem{Lee2016_3Drearrange} W. Lee, H. Kim, and J. Ahn, 
		``Three-dimensional rearrangement of single atoms using actively controlled optical microtraps,''  
		Opt. Express. {\bf24}(9), 9816 (2016).

\bibitem{Barredo2018_3D} D. Barredo, V. Lienhard, S. de. L{\'e}s{\'e}leuc, T. Lahaye, and A. Browaeys, 
		``Synthetic three-dimensional atomic structures assembled atom by atom,'' 
		Nature. {\bf561}, 79-82 (2018).

\bibitem{Kim2020_3DRyd} M. Kim, Y. Song, J. Kim, and J. Ahn, 
		``Quantum Ising Hamiltonian Programming in Trio, Quartet, and Sextet Qubit Systems," 
		PRX Quantum {\bf1}, 020323 (2020).

\bibitem{Song2021_Cayleytree} Y. Song, M. Kim, H. Hwang, W. Lee, and J. Ahn,
``Quantum simulation of Cayley-tree Ising Hamiltonians with three-dimensional Rydberg atoms,''
Phys. Rev. Res. {\bf3}, 013286 (2021).










\bibitem{Byun2023} A. Byun, J. Jung, K. Kim, M. Kim, S. Jeong, H. Jeong, and J. Ahn, 
		``Rydberg-Atom Graphs for Quadratic Unconstrained Binary Optimization Problems,'' 
		 Adv. Quantum Technol. 2300398 (2024).

\bibitem{Liu2017_4B} J. Liu, Y. Qi, Z. Y. Meng, and L. Fu,
		``Self-learning Monte Carlo method,''
		Phys. Rev. B {\bf 95}, 041101(R) (2017).

\bibitem{Dudin2012_superatom} Y. O. Dudin, L. Li, F. Bariani and A. Kuzmich,
		``Observation of coherent many-body Rabi oscillations,"
		Nat. Phys. {\bf8}, 790–794 (2012)

\bibitem{Ebert2015_superatom} M. Ebert, M. Kwon, T. G. Walker, and M. Saffman,
		``Coherence and Rydberg Blockade of Atomic Ensemble Qubits,"
		Phys. Rev. Lett. {\bf115}, 093601(2015).

\bibitem{Zeiher2015_superatom} J. Zeiher, P. Schau{\ss}, S. Hild, T. Macr{\`i}, I. Bloch, and C. Gross
		``Microscopic Characterization of Scalable Coherent Rydberg Superatoms,"
		Phys. Rev. X {\bf5}, 031015 (2015).

\bibitem{Labuhn2016_Ising} H. Labuhn, D. Barredo, S. Ravets, S. de. L{\'e}s{\'e}leuc, T. Macr{\`i}, T Lahaye and A. Browaeys, 
		``Tunable two-dimensional arrays of single Rydberg atoms for realizing quantum Ising models," 
		Nature, {\bf534}, 667-670 (2016).

\bibitem{Derrida1980} B. Derrida,
		``Random-Energy Model: Limit of a Family of Disordered Models,''
		Phys. Rev. Lett {\bf45}, 79 (1980).
\bibitem{Derrida1981} B. Derrida,
		``Random-energy model: An exactly solvable model of disordered systems,''
		Phys. Rev. B {\bf24}, 2613 (1981).

\bibitem{Pichler2018_MIS} H. Pichler, S.-T. Wang, L. Zhou, S. Choi, M. D. Lukin,
		``Quantum Optimization for Maximum Independent Set Using Rydberg Atom Arrays,''
		ArXiv:1808.10816 (2018).  

\bibitem{Zaman_IEEE_2022} M. Zaman, K. Tanahashi and S. Tanaka, ``PyQUBO: Python Library for Mapping Combinatorial Optimization Problems to QUBO Form,'' IEEE Transactions on Computers, {\bf 71}, 4, pp. 838-850 (2022)
\bibitem{Mandal_2020} A. Mandal, A. Roy, S. Upadhyay and H. Ushijima-Mwesigwa, 
				``Compressed Quadratization of Higher Order Binary Optimization Problems,''
					ArXiv:2001.00658 (2020).

\bibitem{DwaveHandbook} D-Wave Systems Inc,
		``D-Wave System Documentation: Problem-Solving Handbook,''
		https://docs.dwavesys.com/docs/latest/doc\_handbook.html (2024).

\bibitem{Boros_DAM_2002} E. Boros and P. L. Hammer ``Pseudo-Boolean optimization,'' Discrete Applied Mathematics {\bf 123}, 1, 155-225 (2002)

\bibitem{Rodrigueze-Heck_thesis_2018} E. Rodr\`iguez-Heck ``Linear ad Quadratic Reformulations of Nonlinear Optimization Problems in Binary Variables,'' PhD Dissertation, Liege University (2018).

\bibitem{Labuhn2014_addressing} H. Labuhn, S. Ravets, D. Barredo, L. B{\'e}guin, F. Nogrette, T. Lahaye, and A. Browaeys, ``Single-atom addressing in microtraps for quantum-state engineering using Rydberg atoms,'' Phys. Rev. A {\bf90}, 023415 (2014).

\bibitem{Omran2019_20addressing} A. Omran, H. Levine, A. Keesling, G. Semeghini, T. T. Wang, S. Ebadi, H. Bernien, A. S. Zibrov, H. Pichler, S. Choi, J. Cui, M. Rossignolo, P. Rembold, S. Montangero, T. Calarco, M. Endres, M. Greiner, V. Vuleti{\'c}, and M. D. Lukin, ``Generation and manipulation of Schr{\"o}dinger cat states in Rydberg atom arrays,'' Science {\bf365}, 570-574 (2019).

\bibitem{Graham2022_MAXCUT}  T. M. Graham, Y. Song, J. Scott, C. Poole, L. Phuttitarn, K. Jooya, P. Eichler, X. Jiang, A. Marra, B. Grinkemeyer, M. Kwon, M. Ebert, J. Cherek, M. T. Lichtman, M. Gillette, J. Gilbert, D. Bowman, T. Ballance, C. Campbell, E. D. Dahl, O. Crawford, N. S. Blunt, B. Rogers, T. Noel, and M. Saffman, ``Multi-qubit entanglement and algorithms on a neutral-atom quantum computer,'' Nature {\bf604}, 457-462 (2022).

\bibitem{deOliveira2024_MWIS}
A. G. de Oliveira, E. Diamond-Hitchcock, D. M. Walker, M. T. Wells-Pestell, G. Pelegr{\'i}, C. J. Picken, G. P. A. Malcolm, A. J. Daley, J. Bass, and J. D. Pritchard
``Demonstration of weighted graph optimization on a Rydberg atom array using local light-shifts,''
ArXiv: 2404.02658 (2024).

\bibitem{Newman1999_ST} M. E. J. Newman and C. Moore
		``Glassy dynamics and aging in an exactly solvable spin model,''
		Phys. Rev. E {\bf60}, 5068 (1999).





\end{thebibliography}
\end{document}